\documentclass[conference]{IEEEtran}
\usepackage{amsmath,amssymb,amsfonts,mathrsfs,graphicx,hhline,subcaption}
\usepackage[ruled,vlined,linesnumbered]{algorithm2e}
\usepackage{geometry}
\usepackage{booktabs} 
\usepackage{CJK}
\usepackage{enumitem,color}

\begin{document}

\title{\emph{CoMap}: Proactive Provision for Crowdsourcing Map in Automotive Edge Computing\vspace{-0.08in}}

\author{
\IEEEauthorblockN{Yongjie Xue, Yuru Zhang, Qiang Liu \\ University of Nebraska-Lincoln \\ qiang.liu@unl.edu\vspace{-0.42in}}
\and
\IEEEauthorblockN{Dawei Chen, Kyungtae Han \\ Toyota InfoTech Labs \\ \{ dawei.chen1, kyungtae.han\}@toyota.com\vspace{-0.42in}}
\vspace{-0.3in}}


\maketitle














\begin{abstract}
Crowdsourcing data from connected and automated vehicles (CAVs) is a cost-efficient way to achieve high-definition maps with up-to-date transient road information.
Achieving the map with deterministic latency performance is, however, challenging due to the unpredictable resource competition and distributional resource demands.
In this paper, we propose \emph{CoMap}, a new crowdsourcing high definition (HD) map to minimize the monetary cost of network resource usage while satisfying the percentile requirement of end-to-end latency.
We design a novel \emph{CROP} algorithm to learn the resource demands of CAV offloading, optimize offloading decisions, and proactively allocate temporal network resources in a fully distributed manner.
In particular, we create a prediction model to estimate the uncertainty of resource demands based on Bayesian neural networks and develop a utilization balancing scheme to resolve the imbalanced resource utilization in individual infrastructures.
We evaluate the performance of \emph{CoMap} with extensive simulations in an automotive edge computing network simulator.
The results show that \emph{CoMap} reduces up to 80.4\% average resource usage as compared to existing solutions.
\end{abstract}


\vspace{-0.05in}
\section{Introduction}
\label{sec:introduction}
Advanced driving assistance systems (ADAS) and autonomous driving will substantially benefit from high-definition (HD) maps in terms of precise relocalization and perception.
HD maps are developed by using a variety of sensors, e.g., camera and LiDAR, to achieve highly accurate representation of road components, e.g., lanes, traffic signs, and interactions.
In general, the data in HD maps are in three categories including stationary data, e.g., roadside buildings, dynamic data~\cite{ahmad2020carmap}, e.g., temporary constructions and accidents, and transient data~\cite{liu2021livemap} such as moving vehicles and pedestrians.
To build HD maps, current strategies mainly count on specialized collection fleets to traverse the road grid, which is cost-inefficient and latency-intolerable for updating the dynamic and transient data in large geographic areas.

Crowdsourcing is the alternative approach, which crowdsources the needed data from connected and automated vehicles (CAVs)~\cite{liu2022edgemap} with a variety of onboard sensors.
By exploiting advanced wireless and edge computing technologies~\cite{liu2019edge}, e.g., 5G and beyond, the crowdsourced data can be transmitted and offloaded to pervasive edge servers for collaborative processing.
To initialize the service of crowdsourcing HD maps, the service provider makes the agreement with the infrastructure providers (e.g., AT\&T and AWS) to obtain the exclusive usage of certain network resources, e.g., wireless bandwidth in base stations and multiple edge servers, for avoiding unpredictable resource competition from other network users, e.g., mobile phones and IoT devices.
Due to the tremendous data size of real-time raw sensor data, e.g., RGB-D images, state-of-the-art works~\cite{ ahmad2020carmap, liu2021livemap,liu2022edgemap} focus on adaptive offloading after partial local processing in CAVs for balancing the network resource usage and computation acceleration.
In particular, the transient data in HD maps need deterministic offloading performance, e.g., $p$th percentile of end-to-end latency.
For example, cooperative perception~\cite{qiu2021autocast} requires that multi-viewed images are captured and processed in very close time stamps.
 

Existing works~\cite{liu2022edgemap, liu2018dare, ran2018deepdecision}, including model-based and model-free approaches, mostly focus on optimizing the average performance for vehicular offloading, e.g., latency and resource usage.
However, we observe that, although the average performance can be achieved, its variance can be substantially large due to the distributional resource demand of individual offloading and changing network dynamics, e.g., radio channels.
In HD maps, the transient data needs to be updated under a consistent latency, e.g., $90$th percentile latency should be less than 100 milliseconds.
The outdated data may provide very limited information regarding current surroundings, e.g., the location of CAVs may change in tens of meters if the latency reaches 500ms or more.
Therefore, it is imperative to investigate new approaches to achieve deterministic performance for crowdsourcing HD maps.

In this paper, we propose \emph{CoMap}, a new crowdsourcing HD map via vehicular offloading from CAVs to edge servers with deterministic latency performance.
The fundamental idea is to optimize the offloading decision and proactively allocate temporal network resources for individual CAV offloading\footnote{The precise resource allocation to particular network users may be accomplished via network slicing techniques~\cite{foukas2016flexran}.}.
The objective is to minimize the monetary cost incurred by temporal network resource usage while satisfying the percentile latency requirement of CAV offloading.
We formulate the optimization problem to seek the optimal offloading decision and temporal radio and computation resource allocation for individual offloading.
To tackle the distributional resource demand of offloading, we create a probabilistic prediction model to learn its uncertainty and generate predictive demands based on Bayesian neural networks.
Then, we design a novel \emph{CROP} algorithm to effectively address the problem under predictive resource demands, in which we design a new utilization balancing scheme to balance the excessive resource demand of CAVs in individual infrastructures.

The contributions of this paper are summarized as follows:
\begin{itemize}
    \item We propose a new crowdsourcing HD map, \emph{CoMap}, via vehicular offloading from CAVs to edge servers with deterministic latency performance.
    \item We design a new probabilistic prediction model to predict the resource demand of individual offloading.
    \item We design a new \emph{CROP} algorithm to minimize the monetary cost while maintaining the percentile latency requirement of offloading. 
    \item We evaluate \emph{CoMap} in an automotive edge computing network simulator, where results show that \emph{CoMap} significantly outperforms existing solutions. 
\end{itemize}


\vspace{-0.05in}
\section{System Overview}

The data plane of \emph{CoMap} is composed of four modules.
The raw RGB-D images captured from CAV sensors are fed into the object detection module, e.g., the YOLO framework, which detects the interested objects, e.g., cars and pedestrians, and generates their bounding boxes.
Next, the feature extraction module (e.g., ORB~\cite{rublee2011orb}) extracts the visual features from the cropped images of detected objects and generates the concise feature representation. 
Then, the extracted features of objects are used to match with the local or global dataset for identifying their historical trajectory. 
Finally, the new updates of \emph{CoMap}, e.g., point cloud and object location, are broadcasted to all CAVs.

The control plane of \emph{CoMap} is composed of three modules.
The prediction module estimates the uncertainty of radio and computation resource demand of offloading in individual CAVs, based on the current observable network states (See Section~\ref{sec:prob_prediction}).
The optimization module determines the offloading decision and temporal resource allocation for individual offloading, based on the estimated uncertainty of resource demands (See Section~\ref{sec:optimization}).
In individual infrastructures, the balancing module adjusts and balances the pre-determined resource allocation of offloading to avoid excessive resource usage (See Section~\ref{sec:balancing}).

\vspace{-0.05in}
\section{System Model}
We consider an automotive edge computing network with multiple cellular base stations (BSs) and edge servers that are distributed in the given geographical area.
Denote $\mathcal{N}$ as the set of CAVs that is connected to the proximal BS and server.
To crowdsource the HD map, all CAVs asynchronously offload their data to the edge server according to the offloading decision $a \in [0, 1]$.
Each vehicular offloading will experience four phases, i.e., local vehicle processing, uplink wireless transmission, edge server computation, and downlink broadcasting.
For instance, the offloading decision $a=0.2$ indicates the first 20\% computation workload will be executed in the CAV and the remaining 80\% computation workload will be processed in the edge server.
As the first 20\% computation completes, the generated intermediate data will be transmitted to its associated edge server via the mobile network.
We denote the set of offloading decisions as $\mathcal{A} = \{ \mathcal{A}^t, \forall t\}$, where $\mathcal{A}^t = \{ a_n^t, \forall n \in \mathcal{N}\}$.
Without loss of generality, we consider that individual CAVs can decide their offloading decision only after the completion of their last offloading to avoid excessive on-the-fly offloading.

\textbf{End-to-End Latency.}
As offloadings span a certain time period to complete multiple sequential phases, it is impossible to derive an exact and precise formulation of end-to-end latency.
This can be attributed to the distributional resource demands of individual offloading and changing network and computing dynamics, e.g., radio channel quality may be varying during the uplink transmission phase.
However, we observe that existing formulations still achieve comparative accuracy under the deterministic resource demands and mild network dynamics~\cite{liu2018dare}.
Denote $f(a_n)$, $g(a_n)$, and $h(a_n)$ as the local computation complexity\footnote{The computation capacity and complexity may be measured by using the metric of GFLOPS and GFLOP, respectively.}, uplink transmission data size, and server computation complexity of the $n$th CAV's offloading.
Then, we formulate the end-to-end latency as 
\begin{equation}
\label{eq:e2e_latency}
L_n = f(a_n) / F_n + g(a_n) / (x_n \cdot E_n) + h(a_n)/ y_n + D_n,
\end{equation}
where $F_n$ and $D_n$ are the computation capacity and static broadcast latency\footnote{In mobile networks, the downlink throughput is usually much greater than that of uplink, which may be attributed to the high transmission power budget of base stations. Considering processed results~\cite{liu2021livemap} are commonly with limited data size, we approximate the broadcasting latency as a static value.}, respectively.
Here, we denote the wireless bandwidth and computation resource allocation as $x_n$ and $y_n$, respectively.
$\mathcal{X} =  \{x_n^t, \forall n, t\}$ and $\mathcal{Y} =  \{y_n^t, \forall n, t\}$ are the collection of all resource allocations in all time slots.
In particular, we introduce $E_n$ as the spectral efficiency of the wireless transmission of the $n$th CAV, which is related to the quality of its radio channel.
Besides, we denote $\mathcal{L} = \{L_n, \forall n\}$ as the collection of end-to-end latency of all offloadings.

\textbf{Problem.} To accomplish real-time \emph{CoMap}, the objective is to minimize the monetary cost incurred by temporal network resource usage, while satisfying the minimum percentile end-to-end latency of offloading in all CAVs.
Given a time period $T$, we formulate the problem $\mathbb{P}_0$ as 
\begin{align}
     \mathbb{P}_0: \min \limits_{\mathcal{A}, \mathcal{X}, \mathcal{Y}} & \;\;\;\;\;  \sum\nolimits_{t = 0}^{T}{\sum\nolimits_{n = 0}^{N} {\left({x_n^t}/{B}+\eta \cdot {y_n^t}/{G}\right)}} \label{eq:org_objective} \\ 
     s.t. & \;\;\;\;\; Pr\left( \mathcal{L } \leq H \right) \ge p, \label{eq:const_latency} \\
     &\;\;\;\;\; 0 \le \sum\nolimits_{n \in \mathcal{N}} x_n^t \le B,   \forall t \label{eq:const_uplink_capacity} \\
     &\;\;\;\;\; 0 \le \sum\nolimits_{n \in \mathcal{N}} y_n^t \le G, \forall t \label{eq:const_server_capacity} \\
     &\;\;\;\;\; 0 \le a_n \le 1,   \forall n, t \label{eq:const_offloading}
\end{align}
where $B$ and $G$ are the total wireless bandwidth and computation capacity of the edge server, respectively.
Besides, $H$ is the latency requirement of individual offloading, and $p \in [0, 1]$ is the required probability of satisfied offloading.
The parameter $\eta$ is introduced to balance the radio and computation resource usage in the objective function of monetary cost.
The constraint in Eq.~\ref{eq:const_uplink_capacity}, Eq.~\ref{eq:const_server_capacity}, and Eq.~\ref{eq:const_offloading} define the optimization space for the uplink radio resource, computation resource and the offloading decision at any time, respectively.
Note that, the resource allocation $x_n^t, y_n^t, \forall t$ of individual offloading are not with a particular time, but span a certain time period.
For example, $x_n^t = 1MHz, \forall t \in [123, 126]$ indicates that there is 1MHz wireless bandwidth reserved for this CAV only from time $t=123$ until $t=126$.

\textbf{Challenges.} 
The problem $\mathbb{P}_0$ is challenging to be resolved in multiple aspects.
First, the problem is with central optimization for all CAVs, which incurs extra communication overhead and delay when transmitting the state of CAVs and deriving the optimal global solution.
In particular, the central optimization usually needs to be executed only for individual offloading, as the other asynchronous offloading may not be finished yet.
Second, the distributional resource demands result in a non-deterministic mathematical expression of the end-to-end latency in Eq.~\ref{eq:e2e_latency}.
As a result, existing off-the-shelf optimization methods, e.g., gradient descents, cannot be applied to solve the problem directly.
Third, the optimization variables span the time period and are highly dimensional, which further complicates the problem-solving consequently.

\vspace{-0.05in}
\section{The CROP Algorithm}
\label{sec:algorithm}
In this section, we propose a new \underline{c}ollaborative dist\underline{r}ibuted \underline{o}ffloading and com\underline{p}utation (CROP) algorithm to effectively resolve the problem $\mathbb{P}_0$. 
First, we reduce the problem into independent offloading problems in individual CAVs in a fully distributed manner, which alleviates the communication overhead and delay in central optimization.
Second, we create a prediction model to learn and predict the distribution of resource demands of individual offloading by using Bayesian neural networks.
Third, we design a new method to derive the optimal offloading decision and temporal resource allocation while maintaining the percentile latency requirement.
Fourth, we design a utilization balancing scheme in individual infrastructures to balance its temporal resource utilization under its instantaneous capacity.





\vspace{-0.05in}
\subsection{Reduced Individualized Problem}
In problem $\mathbb{P}_0$, the correlation among offloadings in CAVs lie in the constraint of resource capacity in Eq.~\ref{eq:const_uplink_capacity} and Eq.~\ref{eq:const_server_capacity}.
Hence, we propose to decouple the problem in terms of offloading in individual CAVs, and we express the reduced problem $\mathbb{P}_1$ in the $n$th CAV as 
\begin{align}
     \mathbb{P}_1: \min \limits_{\mathcal{A}_n, \mathcal{X}_n, \mathcal{Y}_n} & \;\;\;  \sum\nolimits_{t = 0}^{T} ({{x_n^t}/{B}+\eta \cdot {y_n^t}/{G})} \label{eq:org_objective_p1} \\
     s.t. &\;\;\;Pr\left( L_n \leq H \right) \ge p, \label{eq:const_latency_revised} \\
      &\;\;\;\;\; 0 \le x_n^t \le B,   \forall t \label{eq:const_uplink_capacity__} \\
     &\;\;\;\;\; 0 \le y_n^t \le G, \forall t \label{eq:const_server_capacity__} \\
     &\;\;\;\;\; 0 \le a_n \le 1,   \forall n, t \label{eq:const_offloading__}
\end{align}
where we rewrite the constraint in Eq.~(\ref{eq:const_latency_revised}) to assure the percentile latency of offloading in this CAV.
The rationale behind this is that, if we can assure the percentile latency of this CAV's offloading, the requirement $Pr\left( L \leq H \right)$ for all CAVs can also be statistically satisfied.
In addition, we reduce the constraint of resource capacity into Eq.~(\ref{eq:const_uplink_capacity__}) and Eq.~(\ref{eq:const_server_capacity__}), which allows individual CAV to allocate resources independently without the need for other CAVs' information.
The instantaneous constraint of resources will be enforced in the following Sec.~\ref{sec:balancing}. 


\vspace{-0.05in}
\subsection{Probabilistic Demand Prediction}
\label{sec:prob_prediction}
The problem $\mathbb{P}_1$ is non-deterministic as the resource demands ($f, g, h$) are distributional in individual CAV's offloadings.
Hence, we design a new prediction model to learn and predict the resource demands when optimizing the reduced problem $\mathbb{P}_1$ in individual CAVs.

Although the Gaussian process (GP)~\cite{rasmussen2003gaussian} has demonstrated great potential in approximating a variety of black-box functions, its computation complexity $\mathcal{O}(n^3)$, where $n$ is the dimension of collections, leads to poor scalability.
As the offloading of CAVs is usually complete in sub-seconds, the accumulated transitions can reach up to tens of thousands, if not more.
As a result, the training time of the GP model gradually increases under ever-increasing transitions, which fails to achieve real-time decision-making for individual offloading.

Therefore, we design to learn and predict distributional resource demands based on Bayesian neural networks (BNNs), which can scale to accommodate a large number of transitions~\cite{snoek2015scalable}.
Conventional deep neural networks (DNNs) are trained to optimize the fixed neural weights and generate the mean-value predictions by using a variety of loss functions, e.g., cross-entropy and mean square error (MSE).
In contrast, BNNs introduce stochastic components, e.g., neural weights and activation functions, into DNN architectures for quantifying the uncertainty of unknown functions.
For example, the fixed neural weights can be replaced by distributions, which are sampled to be deterministic during the inference.

Different from conventional DNNs, the objective of the BNN training is to find the maximum a posteriori (MAP) weights denoted as $\mathbf{w}^{*} = \arg\underset{\mathbf{w}}{\max} \;{\log P(\mathbf{w}|\mathcal{D})}$.
The neural weights of the BNN are denoted as $\mathbf{w}$ and $\mathcal{D}$ is the accumulative collection of transitions.
According to the Bayes' rule, the calculation of the posterior $P(\mathbf{w}|\mathcal{D})$ requires the prior $P(\mathbf{w})$ and likelihood $P(\mathcal{D}|\mathbf{w})$, which can hardly be practical under large multi-layer DNNs.
Thus, we resort to the variational inference~\cite{blundell2015weight}, which aims to approximate the complex posterior with a simpler and more tractable variational approximation, e.g., Gaussian distribution.
In particular, the posterior $P(\mathbf{w}|\mathcal{D})$ is approximated by minimizing the KL-divergence between the true Bayesian posterior on the weights $KL\left[ q(\mathbf{w}|\theta) || P(\mathbf{w}|\mathcal{D}) \right]$, where $q(\mathbf{w}|\theta)$ is Gaussian distribution with the parameter $\theta$.
Therefore, we can formulate the BNN training that finds the optimal parameter $\theta^*$ 
\begin{align}
    \theta^* = \arg\underset{\theta}{\min}\;{KL\left[ q(\mathbf{w}|\theta)  || P(\mathbf{w})    \right] - \mathbb{E}_{q(\mathbf{w}|\theta)}\left[ \log P(\mathcal{D}|\mathbf{w})\right]}.
\end{align}
Although the minimization of the above function is difficult, if not impossible, we can exploit the \emph{Bayes-by-Backprop}~\cite{blundell2015weight} with the re-parameterization trick to approximate the loss
\begin{equation}
    Loss \approx \sum\limits_{i=1}^{N}{\log q(\mathbf{w}^{i}|\theta) - \log P(\mathbf{w}^{i}) - \log P(\mathcal{D}|\mathbf{w}^{i})},
\end{equation}
where $\mathbf{w}^{i}$ denotes the Monte Carlo sample under the variational posterior $q(\mathbf{w}^{i}|\theta)$.

Then, we create BNNs to learn and predict the resource demands $f, g, h$ based on the experimental measurements.
In particular, we design the state space\footnote{The state space is designed to concisely represent the local observable state from the perspective of individual CAVs. More representative states may be incorporated if applicable.} as the combination of [\emph{offloading decision}, \emph{CAV id}, \emph{CAV location}, \emph{sensor rotation}].
Here, the offloading decision determines the partition of computation and thus significantly affects the resource demands in either CAV or edge server.
The id is the unique identification of the CAV, which helps to identify the particular CAV and potential vehicular properties related to resource demands, e.g., image resolutions.
The CAV location and sensor rotation specify the location and view angle of the sensor, which provides the environmental context, e.g., buildings and walkways.

\vspace{-0.05in}
\subsection{Proactive Predictive Optimization}
\label{sec:optimization}
As a CAV initializes its offloading, the trained BNNs will be invoked to predict the distribution of the resource demands ($f, g, h$).
We observe that the resource demands appear in the numerators of the uplink and computation latency (Eq.~\ref{eq:e2e_latency}), which implies that the percentile of resource demands is directly related to that of the end-to-end latency (Eq.~\ref{eq:const_latency_revised}).
Thus, we propose to convert the constraint in Eq.~\ref{eq:const_latency_revised} from a probabilistic form into a deterministic form as follows.
Given our prediction model generates Gaussian distributions, we calculate the corresponding percentile of the prediction and use percentiles as the deterministic resource demands.
For example, given the predicted mean $\mu$ and std $\sigma$ for resource demands ($f, g, h$), their $90$th and $99$th percentiles are $\mu + 1.28\sigma$ and $\mu + 2.33\sigma$, respectively.
Therefore, the constraint in Eq.~\ref{eq:const_latency_revised} is rewritten as the deterministic $p$th percentile latency, denoted as $\hat{L_n} < H$.
We calculate $\hat{L_n}$ by replacing $f, g, h$ in Eq.~\ref{eq:e2e_latency} with their $p$th percentiles (denoted as $\hat{f}, \hat{g}, \hat{h}$).

Next, we focus on solving the problem $\mathbb{P}_1$ under deterministic resource demands.
As the partition of computation usually with a limited number of discrete values in practical systems~\cite{liu2021livemap}, we propose to exhaustively search for the optimal offloading decision.
Hence, we further reduce the problem $\mathbb{P}_1$ from three kinds of optimization variables into two kinds of continuous variables ($\mathcal{X}_n, \mathcal{Y}_n$).
Then, we solve it by using the Lagrangian primal-dual method~\cite{boyd2004convex}, where Lagrangian is 
\begin{equation}
    \mathcal{L}(x, y, \lambda) = {x_n^t}/{B}+\eta \cdot {y_n^t}/{G} - \lambda \left( \hat{L_n} - H \right),
    \label{eq:lagrangian}
\end{equation}
where the latency constraint is incorporated by introducing a non-negative multiplier $\lambda$.
The problem can be addressed by alternatively solving the primal problem expressed as
\begin{align}
\label{eq:primal_problem}
x^*, y^* & = \arg\min\limits_{x,y \in (\ref{eq:const_uplink_capacity__}), (\ref{eq:const_server_capacity__})}  \mathcal{L}(x, y, \lambda),
\end{align} 
and the dual problem $\lambda^* = \arg\min \limits_{ \lambda \ge 0} \mathcal{L}(x, y, \lambda)$ that can be solved by using the sub-gradient descent~\cite{boyd2004convex}.

We observe that the primal problem in Eq.~\ref{eq:primal_problem} is convex with respect to resource allocations.
Hence, with the Karush-Kuhn-Tucker (KKT) condition, we obtain the optimal resource allocation under the multiplier $\lambda$ as
\begin{align}
     x^* = \sqrt{{(\lambda\cdot \hat{g})}/{(B\cdot E)}}, \;\;\;\;
     y^* = \sqrt{{(\lambda\cdot \hat{h})}/{G}},
\end{align}
where $\hat{g}, \hat{h}$ are the percentile of uplink data size and computation complexity under the given offloading decision.

As the Lagrangian primal-dual method converges, we obtain the optimal resource allocation under different offloading decisions. 
Then, we select the optimal offloading decision with the minimum monetary cost.

Finally, we need to derive the temporal allocation over a time period based on the optimal solution $a^*, x^*, y^*$.
First, we calculate the start and end time of the offloading in all phases, according to the percentile latency $\hat{L_n}$.
Second, we allocate the same radio and computation resource allocation only when the offloading is expected to be wireless transmission and edge computation phase, respectively.

\vspace{-0.05in}
\subsection{Resource Utilization Balancing}
\label{sec:balancing}
In \emph{CoMap}, the asynchronous offloadings usually lead to temporal overlaps in different phases.
As resource allocations are independently optimized in individual CAVs, the aggregated resources may exceed the overall capacity in individual infrastructures, i.e., total wireless bandwidth $B$ and computation capacity $G$.
The excessive resource allocations cannot be fulfilled by infrastructures, which results in delays in completing these offloadings.

To address this issue, we design a utilization balancing scheme in individual infrastructures to balance the temporal resource utilization among all CAVs. 
This is based on the observation that the offloading can always be completed as long as the accumulated resource allocation exceeds its computation complexity.
For instance, given the computation resource allocation is $y_n^t = 1, t \in [123, 126]$, the $n$th CAV's offloading may be completed in a single time slot, e.g., $t=125$ if $y_n^{125} = 4$.

Specifically, as every resource allocation is received by the infrastructure, we balance the resource utilization as follows.
First, we calculate the average utilization, including the new resource allocation, in the time period of this offloading.
Second, we remove the time slots whose allocated resources exceed the average utilization, because these time slots are already saturated.
Third, we re-calculate the average resource utilization on the remaining time slots of this offloading.
Fourth, we deduct the already allocated resources from the average resource utilization, and the outcome will be clipped to be non-negative and selected as the allocated resource for this offloading.
This method is inspired by the water-filling algorithm to balance the utilization of resources.

\begin{figure*}[!t] 
\captionsetup{justification=centering}
  \begin{minipage}[t]{0.325\textwidth}
    \centering
    \includegraphics[width=2.3in]{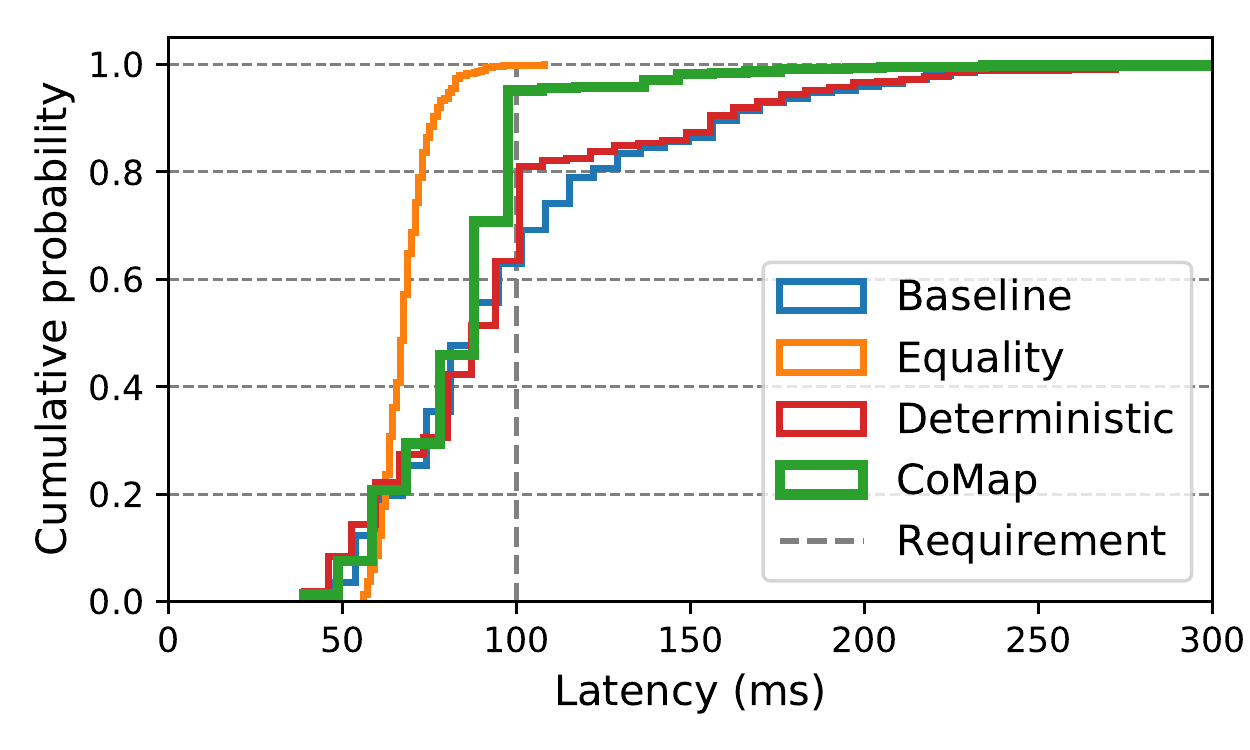}
    \captionof{figure}{\small Latency performance of algorithms}
    \label{fig:results_offload_latency}
  \end{minipage}
  \hfill
  \begin{minipage}[t]{0.325\textwidth}
    \centering
    \includegraphics[width=2.3in]{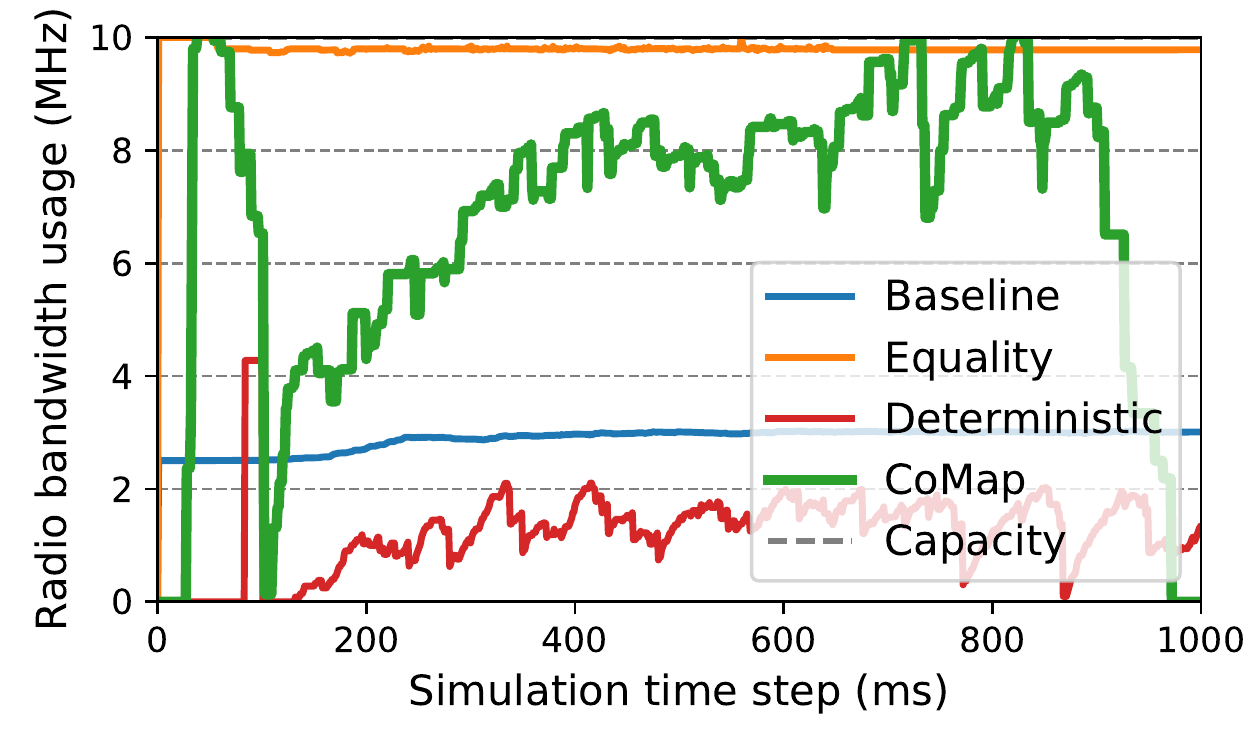}
    \captionof{figure}{\small Radio resource allocation}
    \label{fig:results_usage_radio}
  \end{minipage}
  \hfill
  \begin{minipage}[t]{0.325\textwidth}
    \centering
    \includegraphics[width=2.3in]{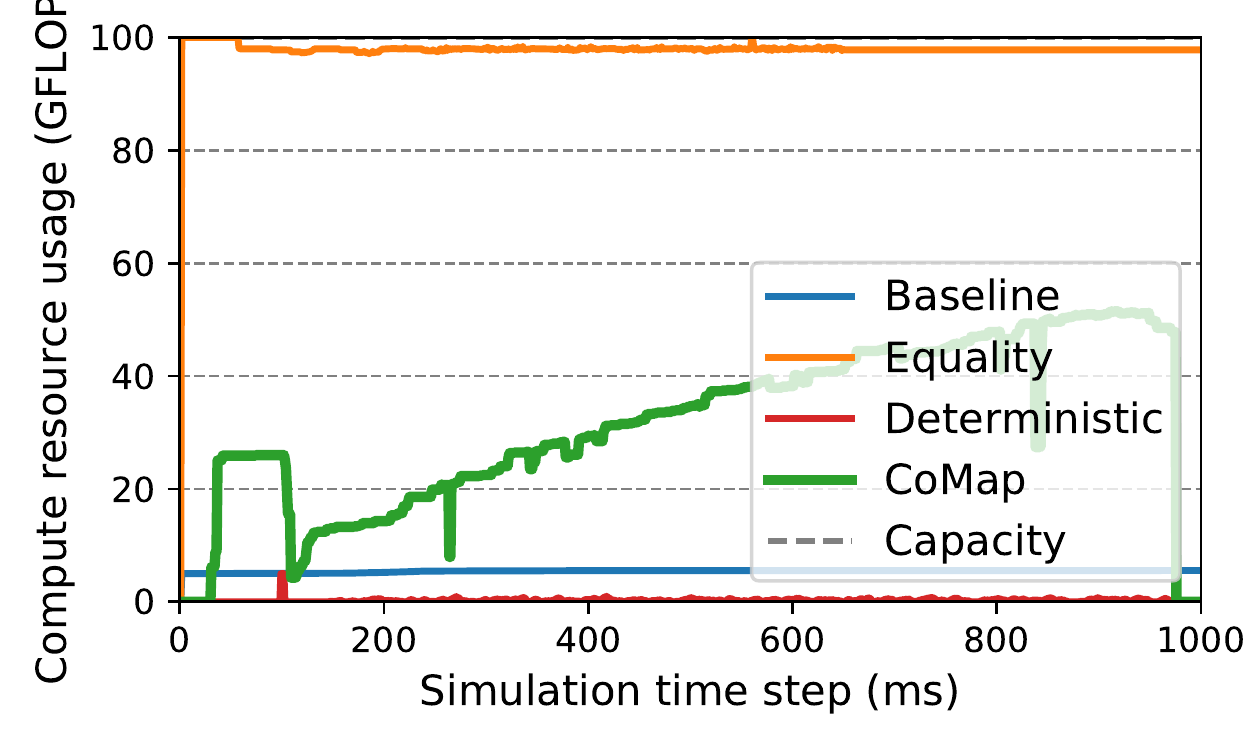}
    \captionof{figure}{\small Compute resource allocation}
    \label{fig:results_usage_compute}
  \end{minipage}
\end{figure*}

\begin{figure*}[!t] 
\captionsetup{justification=centering}
  \begin{minipage}[t]{0.325\textwidth}
    \centering
    \includegraphics[width=2.3in]{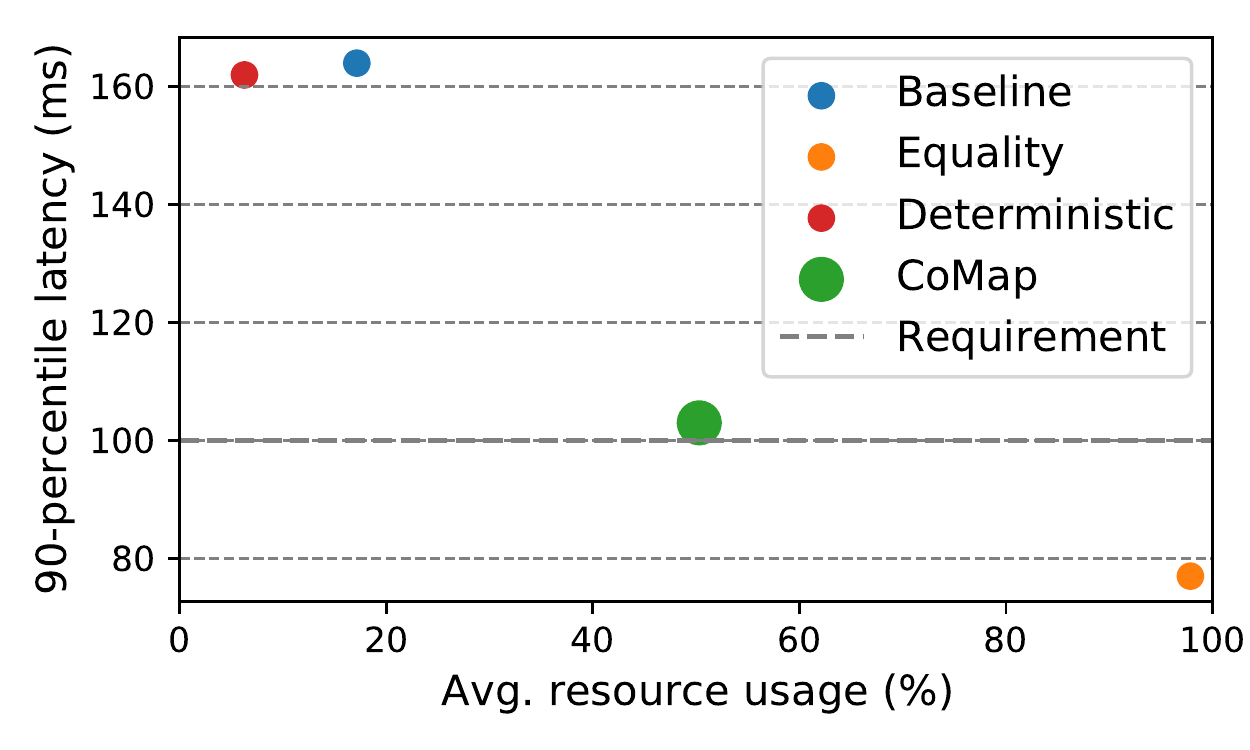}
    \captionof{figure}{\small Performance of algorithms}
    \label{fig:results_scatter}
  \end{minipage}
  \hfill
  \begin{minipage}[t]{0.325\textwidth}
    \centering
    \includegraphics[width=2.3in]{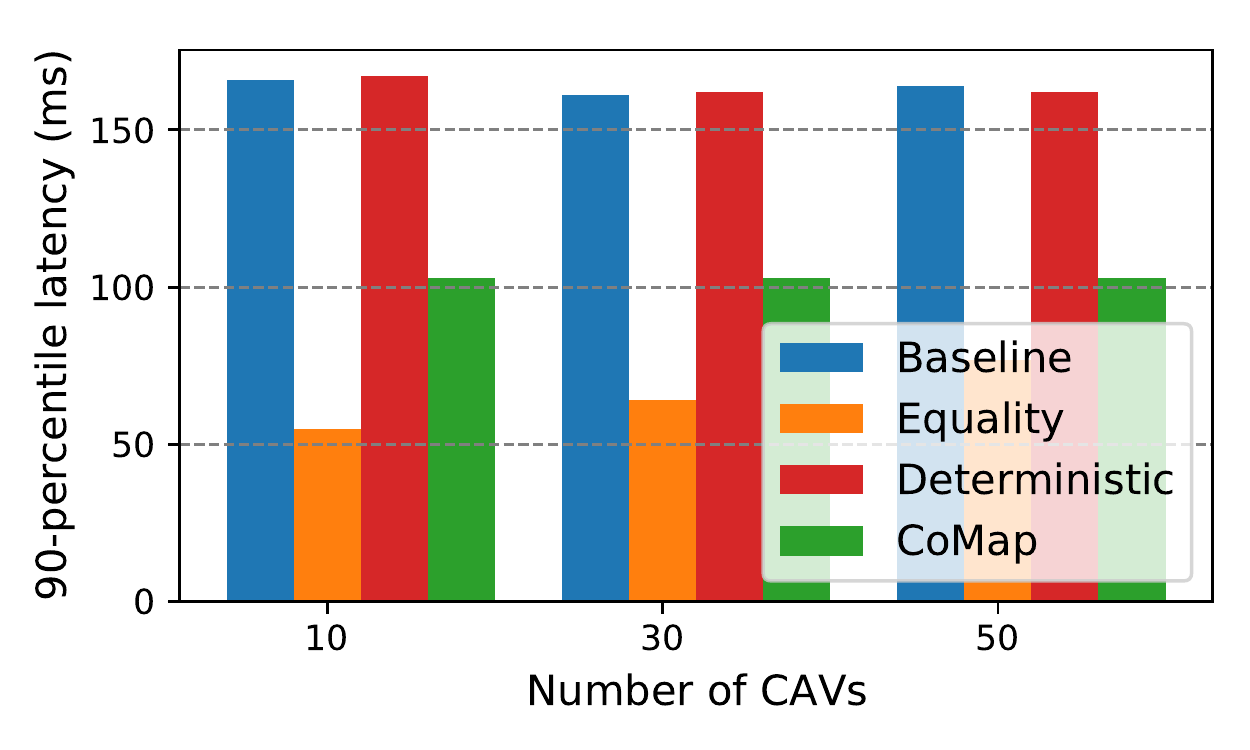}
    \captionof{figure}{\small Latency under different traffic}
    \label{fig:results_traffic_latency}
  \end{minipage}
  \hfill
  \begin{minipage}[t]{0.325\textwidth}
    \centering
    \includegraphics[width=2.3in]{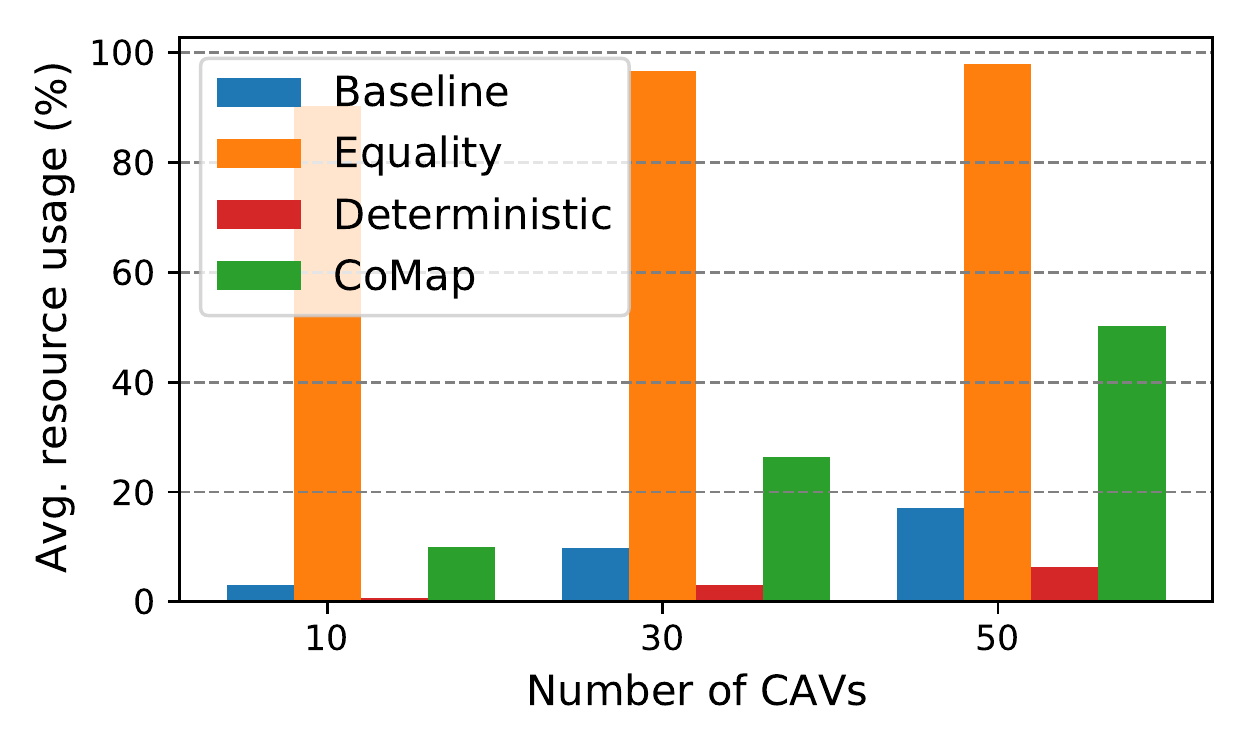}
    \captionof{figure}{\small Usage under different traffic}
    \label{fig:results_traffic_usage}
  \end{minipage}
\end{figure*}

\vspace{-0.05in}
\section{Performance Evaluation}
\label{sec:evaluation}

\textbf{Network Simulator.}
We develop an automotive edge computing network simulator including the component of a server computation, an uplink wireless transmission, a downlink broadcasting, and vehicular computations for individual CAVs.
The simulator is designed to be time-driven, in other words, all the components are looped and their parameters are updated every simulation time step.

To simulate the end-to-end communication and computation of CAV offloading, we create a \emph{task} class in Python and traverse \emph{tasks} in these consequential components.
The offloading is completed only if its \emph{task} finishes in the downlink broadcast component.
A \emph{task} includes all the offloading parameters, such as remaining local and server computation complexity, and the remaining wireless transmission size.
The offloading parameters are determined via sampling from the experimental measurements.
We adopt a 5G simulator~\cite{oughton2019open} in the wireless transmission component, where the radio channel is urban micro (UMi - Street Canyon) and the trajectory of CAVs are based on the V2X-Sim dataset~\cite{Li_2021_RAL}.
We build the vehicular and server computation component by using a single FIFO service queue.
In each simulation time step (e.g., 1 ms), the parameters in \emph{tasks} are updated according to the allocated resources in different components.
For example, the achievable data rate of CAVs is calculated in the wireless communication module, and the remaining transmission size is deducted accordingly. 
If there are no allocated resources in a time step, a minimum resource will be assigned, e.g., 50kHz wireless bandwidth and 0.1 GFLOPS computation capacity.

\textbf{Parameters.}
We conduct extensive experiments to profile the resource demands of vehicular offloading in \emph{CoMap}.
We adopt the V2X-Sim dataset~\cite{Li_2021_RAL}, which includes 100 CAVs at 100 frames with a variety of sensor data such as RGB image, depth, and LiDAR.
The dataset is obtained via CARLA-SUMO co-simulation in the default scenarios in the CARLA simulator.
We develop the data plane of \emph{CoMap} with YOLOv5m object detector, ORB feature extraction, and brutal-force feature matching.
The offloading decision can be selected from [0.0, 0.33, 0.66, 1.0], which correspond to the partition after the raw data retrieval, object detection, feature extraction, and feature matching, respectively.
The profiling is conducted on an Intel i7 desktop with 16G RAM and 1TB M.2. SSD.
Without loss of generality, we consider this desktop has the same computation capacity as CAVs, and the edge computation capacity is 100x than that of CAVs\footnote{We aim to reduce the monetary cost incurred by resource usage, the large computation capacity of edge server is mainly to accommodate more CAVs, rather than achieving 100x acceleration for individual offloadings.}.
The average transmission data size are $g(0)=992.06Kbits$, $g(0.33)=337.43Kbits$, $g(0.66)=9.56Kbits$, and $g(1)=9.56Kbits$.
The average local computation complexity are $h(0)=0GFLOP$, $h(0.33)=3.74GFLOP$, $h(0.66)=4.10GFLOP$, and $h(1)=8.21GFLOP$, where the computation capacity of CAV $F$ is normalized as 1GFLOPS, for the sake of simplicity.
Other parameters are listed as $\eta=1$, $B=10MHz$, $E=8bps/Hz$, $H=100ms$, and $p=90$th percentile.
The default number of CAVs is 50.

We compare \emph{CoMap} with the following algorithms:
\begin{itemize}
    \item \emph{Baseline}: The \emph{Baseline} completes all the computation components onboard in CAVs, while using the minimum resource allocation for individual offloading throughout the following one second.
    \item \emph{Equality}: The \emph{Equality} shares all network resources equally to all CAVs for all time slots, where offloading decisions are exhaustively searched to minimize the percentile latency requirement.
    \item \emph{Deterministic}: The \emph{Deterministic}~\cite{liu2018dare} regresses the distributional resource demands via polynomial regression with respect to offloading decisions. It generates the mean-value prediction of resource demands, which are used to optimize the offloading decision and resource allocations accordingly.
\end{itemize}

\textbf{Latency Performance.} 
Fig.~\ref{fig:results_offload_latency} shows the cumulative probability of end-to-end latency of vehicular offloading under different algorithms.
We observe that \emph{CoMap} and \emph{Equality} meet the probabilistic latency requirement $P(\mathcal{L}<100ms)>=0.9$, while \emph{Baseline} and \emph{Deterministic} fail. 
Note that there are nearly 20\% latencies located in the range of $[95ms, 100ms]$ in \emph{CoMap}, which is intentionally optimized in the \emph{CROP} algorithm to meet the latency with the minimum resource usage.
As \emph{Deterministic} relies on the regressed model with only mean value predictions, the percentile latency requirement cannot be effectively satisfied.
This result justifies the necessity of probabilistic prediction of resource demands in optimizing resource allocations under changing network dynamics.

\textbf{Utilization Performance.} 
\emph{Equality} achieves the best latency performance in all algorithms, which is attributed to its very high resource utilization as shown in Fig.~\ref{fig:results_usage_radio} and Fig.~\ref{fig:results_usage_compute}.
In particular, we observe that \emph{Deterministic} tends to choose the offloading decision as 1, which completes all the components in CAVs and requires fewer resources in both wireless communication and edge computation.
Fig.~\ref{fig:results_scatter} is the scatter plot of the average resource usage and $90$th percentile latency achieved by all algorithms.
We can see that \emph{CoMap} obtains the best overall performance, i.e., reducing the monetary cost incurred by resource usage while almost exactly matching the percentile latency requirement.

\textbf{Scalability Performance.} 
Fig.~\ref{fig:results_traffic_latency} and Fig.~\ref{fig:results_traffic_usage} show the latency and resource usage of algorithms under different numbers of CAVs.
As more CAVs are in the network, the overlaps among vehicular offloadings become more intensive and the potential of excessive resource allocation may be more frequent.
Hence, we observe that the $90$th percentile latency performance of most algorithms are deteriorated, even if their resource usage is increased.
In contrast, \emph{CoMap} can maintain its latency performance with the increased average usage of network resources.
In particular, \emph{CoMap} reduces 70.2\% and 80.4\% average resource usage than \emph{Equality} under 10 and 30 CAVs, respectively.
This result validates the scalability of \emph{CoMap} to tackle changing network traffic dynamics.

\begin{figure}[!t]
	\centering
	\includegraphics[width=2.3in, height=1.2in]{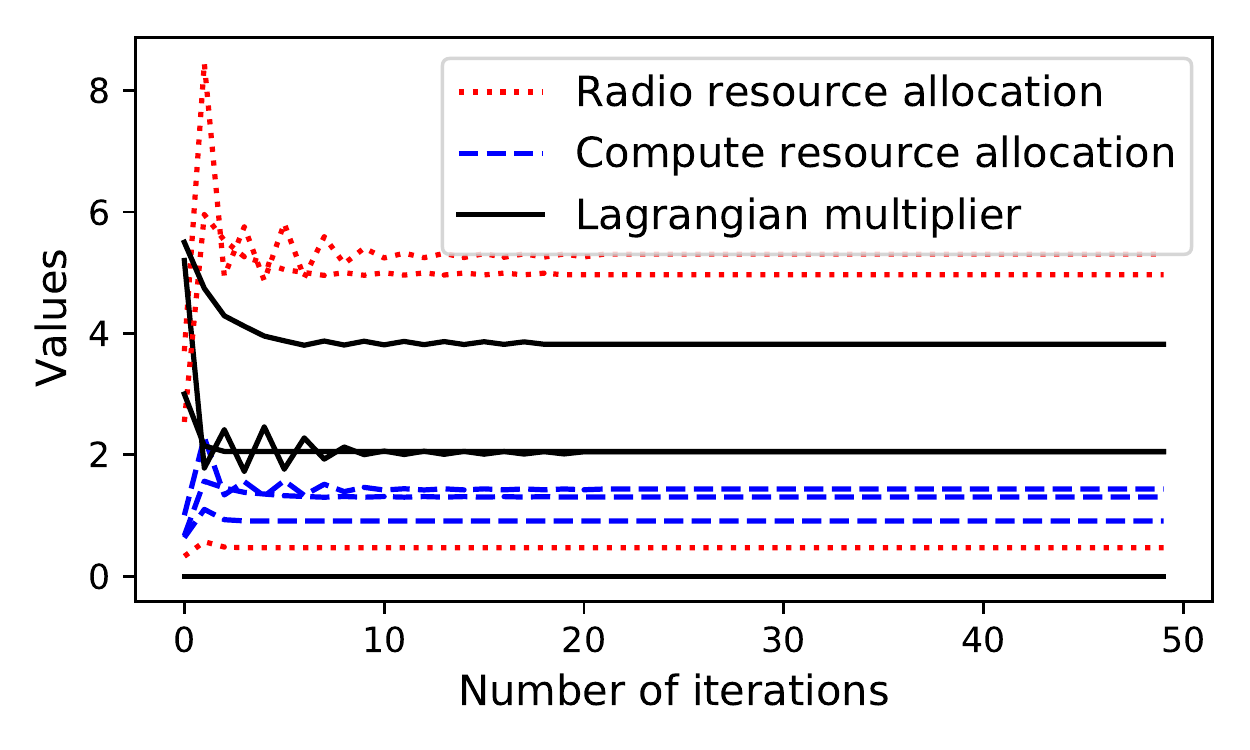}
	\vspace{-0.05in} \caption{Convergence of \emph{CoMap}}
	\label{fig:results_convergence}
\end{figure}

\textbf{Convergence Performance.}
In Fig.~\ref{fig:results_convergence}, we show an exemplary snapshot regarding the convergence of \emph{CoMap}.
In the \emph{CROP} algorithm, it uses the Lagrangian primal-dual method to iteratively search for the optimal allocation of radio and computation resources.
We see that the \emph{CROP} algorithm converges in several iterations, where the duplicated curves are obtained from different offloading decisions.
This result shows that the \emph{CROP} algorithm achieves a fast convergence rate, which maintains low computation complexity in executing the algorithm at runtime in individual CAVs.

\vspace{-0.05in}
\section{Related Work}
Computation offloading~\cite{ liu2018joint, zhang2018joint} has been extensively investigated to exploit the powerful edge server to accelerate the computation of mobile devices, e.g., smartphones and vehicles.
The optimization problems are formulated under a variety of network settings, e.g., snapshot-based~\cite{liu2018joint, zhao2019computation} and Markov decision process~\cite{liu2019deep}, and diverse objective functions, e.g., latency, energy consumption, and system costs.
For instance, Zhao \emph{et. al}~\cite{zhao2019computation} optimized the computation offloading and resource allocation for achieving Nash equilibrium (NE) via game theory.
Existing solutions can be generally categorized into model-based (e.g., convex optimization~\cite{zhang2018joint}) and model-free (e.g., deep reinforcement learning~\cite{chen2021drl}).
For example, Liu \emph{et. al}~\cite{liu2019deep} proposed a Q-learning and a DRL-based method to address the offloading problem, which maximizes the long-term communication and computation utility under the constraint of task latency.
While most existing works focus on optimizing the average performance, we aim to satisfy the percentile latency requirements.
Besides, we are the first to proactively allocate network resources to vehicular offloading, where the resource allocations are temporal and span certain time periods.

\vspace{-0.05in}
\section{Conclusion}
In this paper, we designed a new crowdsourcing HD map in automotive edge computing. We proposed the \emph{CROP} algorithm to optimize \emph{CoMap} by determining the optimal vehicular offloading decision and temporal radio and computation resource allocations. We evaluated the performance of \emph{CoMap} via extensive network simulations, where the results validate the efficacy of \emph{CoMap} over existing solutions.

\section*{Acknowledgement}
This work is partially supported by the US National Science Foundation under Grant No. 2212050.

\vspace{-0.05in}
\bibliographystyle{IEEEtran}
\bibliography{ref/reference, ref/qiang}

\begin{thebibliography}{10}
\providecommand{\url}[1]{#1}
\csname url@samestyle\endcsname
\providecommand{\newblock}{\relax}
\providecommand{\bibinfo}[2]{#2}
\providecommand{\BIBentrySTDinterwordspacing}{\spaceskip=0pt\relax}
\providecommand{\BIBentryALTinterwordstretchfactor}{4}
\providecommand{\BIBentryALTinterwordspacing}{\spaceskip=\fontdimen2\font plus
\BIBentryALTinterwordstretchfactor\fontdimen3\font minus
  \fontdimen4\font\relax}
\providecommand{\BIBforeignlanguage}[2]{{%
\expandafter\ifx\csname l@#1\endcsname\relax
\typeout{** WARNING: IEEEtran.bst: No hyphenation pattern has been}%
\typeout{** loaded for the language `#1'. Using the pattern for}%
\typeout{** the default language instead.}%
\else
\language=\csname l@#1\endcsname
\fi
#2}}
\providecommand{\BIBdecl}{\relax}
\BIBdecl

\bibitem{ahmad2020carmap}
F.~Ahmad \emph{et~al.}, ``Carmap: Fast 3d feature map updates for
  automobiles,'' in \emph{17th $\{$USENIX$\}$ Symposium on Networked Systems
  Design and Implementation ($\{$NSDI$\}$ 20)}, 2020, pp. 1063--1081.

\bibitem{liu2021livemap}
Q.~Liu, T.~Han, J.~L. Xie, and B.~Kim, ``Livemap: Real-time dynamic map in
  automotive edge computing,'' in \emph{IEEE INFOCOM 2021-IEEE Conference on
  Computer Communications}.\hskip 1em plus 0.5em minus 0.4em\relax IEEE, 2021,
  pp. 1--10.

\bibitem{liu2022edgemap}
Q.~Liu, Y.~Zhang, and H.~Wang, ``Edgemap: Crowdsourcing high definition map in
  automotive edge computing,'' in \emph{IEEE International Conference on
  Communications}.\hskip 1em plus 0.5em minus 0.4em\relax IEEE, 2022, pp.
  4300--4305.

\bibitem{liu2019edge}
S.~Liu, L.~Liu, J.~Tang, B.~Yu, Y.~Wang, and W.~Shi, ``Edge computing for
  autonomous driving: Opportunities and challenges,'' \emph{Proceedings of the
  IEEE}, vol. 107, no.~8, pp. 1697--1716, 2019.

\bibitem{liu2018dare}
Q.~Liu and T.~Han, ``Dare: Dynamic adaptive mobile augmented reality with edge
  computing,'' in \emph{2018 IEEE 26th International Conference on Network
  Protocols (ICNP)}.\hskip 1em plus 0.5em minus 0.4em\relax IEEE, 2018, pp.
  1--11.

\bibitem{ran2018deepdecision}
X.~Ran \emph{et~al.}, ``Deepdecision: A mobile deep learning framework for edge
  video analytics,'' in \emph{IEEE Conference on Computer
  Communications}.\hskip 1em plus 0.5em minus 0.4em\relax IEEE, 2018, pp.
  1421--1429.

\bibitem{foukas2016flexran}
X.~Foukas \emph{et~al.}, ``{FlexRAN}: A flexible and programmable platform for
  software-defined radio access networks,'' in \emph{ACM CoNEXT}, 2016, pp.
  427--441.

\bibitem{rublee2011orb}
E.~Rublee \emph{et~al.}, ``Orb: An efficient alternative to sift or surf,'' in
  \emph{2011 International conference on computer vision}.\hskip 1em plus 0.5em
  minus 0.4em\relax IEEE, pp. 2564--2571.

\bibitem{rasmussen2003gaussian}
C.~E. Rasmussen, ``Gaussian processes in machine learning,'' in \emph{Summer
  school on machine learning}.\hskip 1em plus 0.5em minus 0.4em\relax Springer,
  2003, pp. 63--71.

\bibitem{snoek2015scalable}
J.~Snoek \emph{et~al.}, ``Scalable bayesian optimization using deep neural
  networks,'' in \emph{International conference on machine learning}.\hskip 1em
  plus 0.5em minus 0.4em\relax PMLR, 2015, pp. 2171--2180.

\bibitem{blundell2015weight}
C.~Blundell \emph{et~al.}, ``Weight uncertainty in neural network,'' in
  \emph{International conference on machine learning}.\hskip 1em plus 0.5em
  minus 0.4em\relax PMLR, 2015, pp. 1613--1622.

\bibitem{boyd2004convex}
S.~Boyd, S.~P. Boyd, and L.~Vandenberghe, \emph{Convex optimization}.\hskip 1em
  plus 0.5em minus 0.4em\relax Cambridge university press, 2004.

\bibitem{oughton2019open}
E.~J. Oughton, K.~Katsaros, F.~Entezami, D.~Kaleshi, and J.~Crowcroft, ``An
  open-source techno-economic assessment framework for {5G} deployment,''
  \emph{IEEE Access}, vol.~7, pp. 155\,930--155\,940, 2019.

\bibitem{Li_2021_RAL}
Y.~Li, D.~Ma, Z.~An, Z.~Wang, Y.~Zhong, S.~Chen, and C.~Feng, ``V2x-sim:
  Multi-agent collaborative perception dataset and benchmark for autonomous
  driving,'' \emph{IEEE Robotics and Automation Letters}, 2022.

\bibitem{liu2018joint}
Q.~Liu, T.~Han, and N.~Ansari, ``Joint radio and computation resource
  management for low latency mobile edge computing,'' in \emph{IEEE Global
  Communications Conference (GLOBECOM)}.\hskip 1em plus 0.5em minus 0.4em\relax
  IEEE, 2018, pp. 1--7.

\bibitem{zhang2018joint}
J.~Zhang \emph{et~al.}, ``Joint computation offloading and resource allocation
  optimization in heterogeneous networks with mobile edge computing,''
  \emph{IEEE Access}, vol.~6, pp. 19\,324--19\,337, 2018.

\bibitem{zhao2019computation}
J.~Zhao \emph{et~al.}, ``Computation offloading and resource allocation for
  cloud assisted mobile edge computing in vehicular networks,'' \emph{IEEE
  Trans. on Vehicular Technology}, vol.~68, no.~8, pp. 7944--7956, 2019.

\bibitem{liu2019deep}
Y.~Liu \emph{et~al.}, ``Deep reinforcement learning for offloading and resource
  allocation in vehicle edge computing and networks,'' \emph{IEEE Transactions
  on Vehicular Technology}, vol.~68, no.~11, pp. 11\,158--11\,168, 2019.

\bibitem{chen2021drl}
J.~Chen \emph{et~al.}, ``A drl agent for jointly optimizing computation
  offloading and resource allocation in mec,'' \emph{IEEE Internet of Things
  Journal}, 2021.

\end{thebibliography}

\end{document}